\documentstyle[aps,prb,epsfig,graphicx]{revtex}
\long\def\title#1{{\Large\begin{center}#1\end{center}\par}}
\long\def\address#1{\begin{center}#1\end{center}\par}
\long\def\author#1{\begin{center}#1\end{center}\par}
\def\pacs{}
\tightenlines

\begin{document}

\title{Phonon softening: ``keyhole'' into dynamic stripe-phase}

\author{Sergei I. Mukhin}
\address{Theoretical Physics Department,
Moscow Institute for Steel and Alloys, Leninskii pr. 4, 119991 Moscow,
Russia}

\vspace{3mm}Running title: soft phonons and dynamic stripes 

\vspace{3mm} Keywords: high-T$_c$ cuprates, in-plane optical phonon modes,
phonon self-energy, meandering stripes, on-stripe spinless fermionic holes 
\vspace{3mm}
\draft

\begin{abstract}
The highly distinctive phonon self-energy dependences on the wave vector, 
calculated respectively for the static and dynamic stripe phase models of the 
underdoped cuprates are presented. The negative values of the real part of the
on-stripe holes 
polarization loop lead to appearance of localized vibration states in the 
presence of well-separated stripes, relevant for underdoped cuprates. The
localized modes split below the bare optical phonon frequencies
exhibiting strong ``softening effect''. 
The calculated gap between localized and propagating phonon frequencies is
practically doping independent in accord with the recent neutron
measurements. The derived wave vector dependence of the phonon softening 
nicely coincides with anomalously shallow experimental curve in
high-T$_c$ cuprates when stripe meandering is taken into account. 
Co-centering of the theoretical and experimental curves implicates on-stripe 
holes statistics.  
\end{abstract}


\section{Introduction and main results}
\begin{figure}[tbph]
\begin{center} 
\includegraphics[width=3.0in]{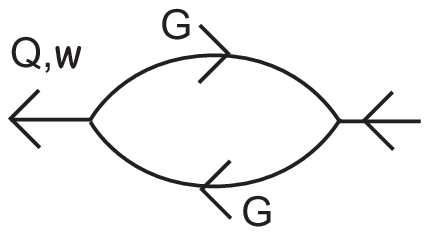}
\caption{Phonon self-energy $\pi(Q,\omega)$ : single-loop diagram for 1D
on-stripe holes. Dot is electron(hole)-lattice coupling
vertex, the  thin lines are Green's functions $G(p,\epsilon)$ of the 1D hole
possesing momentum $p$ along the stripe direction.}
\end{center}
\label{loopa}
\end{figure}

The independence on doping of the amount of softening of the optical phonon
modes measured by neutrons in high-Tc cuprates \cite{Qee,pintschovius},
concomitant with the linear doping-dependence of the modes intensity, points 
to a local on-stripe character of the softening effect \cite{alan}. The latter
effect may arise due to coupling of the in-plane optical Cu-O stertching modes 
(LO) to the holes populating stripes. In this paper, in order to explore the 
effect theoretically, the stripes are considered as one-dimensional (1D)
``metallic rivers of doped holes" \cite{Kivelson}. Spacially localized
vibrations are found, that split below the bare phonon band in the 
stripe-phase,  as was proposed earlier \cite{sm}. The source of 
localization of the lattice 
vibrations is the  phonon self-energy, significantly space-modulated in 
the presence of the on-stripe holes. It is shown that this self-energy plays a
role of attractive localizing potential for the phonon quasi-particles, 
provided that its real part is negative and has sufficiently high absolute 
value. The latter depends on the wave-vector in the direction along the 
stripes. In this direction the translational symmetry of the crystal lattice 
is preserved.
It is demonstrated here that phonon self-energy, calculated from the standard 
Feynman diagram Fig.\ref{loopa}, cannot lead to the peculiar flat 
momentum dependence of the optical mode softening measured by neutron 
scattering in La-based high-T$_c$ cuprates, see filled circles in 
Fig.\ref{comparison}. As it is
apparent from Fig.\ref{comparison}, a rather poor agreement with experimental 
data is obtained for the case of the holes moving along 1D static stripes 
(dash-dotted line), as well as for the case of the holes moving 
three(two)-dimensionally (dotted line). Miraculously, inclusion into phonon 
self-energy of the overdamped long wavelength stripe-meandering propagator, 
see Fig. \ref{loopb}, leads to impressive coincidence of the theoretical 
phonon softening curve (solid line) in  Fig. \ref{comparison}, with the
experimental data  (filled circles). Besides the flat momentum dependence of
the modes softening, the co-centering of the experimental and
theoretical curves is also important. Position of zero along the $Q$-axis 
of the theoretical curve depends on the Fermi momentum of the on-stripe holes. 
This leads to important observation that the theoretical phonon dispersion 
curve coincides with the experimental one, provided that the underlying 1D 
``Fermi momentum'' along stripes is twice the value predicted in the 
weak-coupling scheme. This discrepancy may be tolerated if one assumes the 
holes inside stripes to be spinless. Incidentally, the energy-integrated 
ARPES data \cite{shen} exhibit flat parallel pieces enclosing high spectral 
weight region in the Brillouin zone (the ``holy cross'' picture), that 
resemble 1D ``Fermi surface'' of the on-stripe holes, but with Fermi momentum 
value corresponding to the weak-coupling scheme. This may indicate that 
``spin-dressed'' holes rather than spinless are involved in the ARPES 
measurements. 

Section II contains derivation of the  dynamic elasticity equations for the 2D 
plane with periodic stripes. Section III uses straightforward mapping of the
elasticity equations onto Schr\"odinger equations for a quantum particle in
the periodic  Dirac's haircomb potential. Localized solutions describe
``softened phonon'' modes. In Sections IV and V different dependences of
the phonon 
self-energy on the wave vector along the stripes are
calculated for static and meandering 1D stripes respectively. Section VI 
contains description of the model parameters found by fitting the theoretical 
curves to experimental data, and concludes with discussion of 
the main results briefly reviewed above. Some lengthy analytical 
expressions are located in the Appendices \ref{App2} and \ref{App1}.

\section{2D lattice dynamics in the stripe-phase}
Elastic behavior of 2D quadratic lattice (modeling CuO planes in cuprates)
is considered in the continuum media approximation.
Corresponding expression for the elastic
energy density reads (see e.g. \cite{Kleinert}):

\begin{eqnarray} 
\varepsilon(\vec{r})=\frac{1}{2}(2\mu+\lambda)\displaystyle(u_{11}^2+u_{22}^2)+
\lambda u_{11}u_{22}+2\mu u_{12}^2.
\label{energy}
\end{eqnarray}

\noindent
here $\lambda$ is Lam\'e constant; $\mu$ is shear modulus. 
The strain tensor components $u_{ij}=$, $i,j=1,2$ are : 

\begin{eqnarray}
u_{ij}=\displaystyle\frac{1}{2}(\partial_iu_j+\partial_ju_i+
\partial_iu_l\partial_ju_l)
\label{strain}
\end{eqnarray}

\noindent
where $u_x,\;u_y$ are the in-plane local displacements 
of the lattice; $\partial_i\equiv \partial/\partial x_i$ and coordinates $x_i$ 
belong to the plane $\{x,y\}$.

\noindent 
Assume that stripes are oriented along $y$-axis in the $\{x,y\}$ plane. Hence, 
in the static picture they form a periodic array with a period $b\sim 1/x_h$ 
along the
$x$-axis, where $x_h$ is the concentration of doped holes in the Cu-O plane
counted from the insulating state $x_h=0$. Then, translation invariance along
$y$-axis is preserved and therefore vibrations keep $y$-component of
quasi momentum $Q$.
Corresponding dynamic elasticity 
equations in the mixed momentum-coordinate representation take the form:

\begin{eqnarray}
&&\displaystyle \left[\rho\omega^2-\mu Q^2+(\mu+K)\partial^2_{xx}-
\pi_1(Q,\omega)p(x)\right]u_x+
iKq\partial_x u_y=0\nonumber\\
&&\displaystyle \left[\rho\omega^2-(\mu+K)Q^2+\mu\partial^2_{xx}-
\pi_2(Q,\omega)p(x)\right]u_y+
iKq\partial_x u_x=0                            
\label{dynamo}
\end{eqnarray}

\noindent where $\rho$ is the lattice density of mass and $K=\lambda+\mu$ is
the lattice compression modulus in the absence of doped holes. Here $\omega$ 
is brought by the Fourier transformation along the time axis $t$:

\begin{eqnarray}
 u_x=u_x(x)\exp\{iQy-i\omega t\}; u_y=u_y(x)\exp\{iQy-i\omega t\},
\label{vibra}
\end{eqnarray}

\noindent
The $Q,\omega,x$-dependent phonon self-energy is included in the
dynamics equations (\ref{dynamo}) in the simplest possible way via the terms
$\pi_i(Q,\omega)p(x)$. Here $Q,\omega$-dependence is due to hole motion along
stripes inside the ``metallic rivers of charge'' separating antiphased
antiferromagnetic domains in the ``canonical'' stripe phase pattern 
\cite{Zaanen,Kivelson}. A model simplifies the general tensorial self-energy 
$\pi_{ij}(Q,\omega,x)$ with a diagonal matrix, maintaining essential physics 
of phonon localization, but avoiding intractable algebra:

\begin{eqnarray}
\pi_{xy}=0;\;\pi_{xx}=\pi_1(Q,\omega)p(x),\;\pi_{yy}=\pi_2(Q,\omega)p(x);
\label{symms}
\end{eqnarray}

\noindent where a ``tight binding'' approximation is used to mimic periodic 
$x$-dependence in the direction perpendicular to the stripes:

\begin{eqnarray}
p(x)=\sum_n\delta(x+bn)
\label{dirac}
\end{eqnarray}

\noindent Here $n$ enumerates the stripes and $b$ is the interstripe period.
This approximation looks reasonable for underdoped cuprates, since according 
to e.g. \cite{tranq} the number of doped holes/per double-period $2a$ along 
a stripe 
is fixed to one, and therefore the interstripe period $b=a/2x_h\gg a$ in the 
low hole-doping limit $x_h\ll 1$ for a square lattice. In the limit of 
vanishing
electron-phonon coupling $\pi_{i}(Q,\omega)\rightarrow 0$, the system 
(\ref{dynamo}) smoothly
transforms into a standard system of dynamic equations for 2D isotropic
elastic media that possesses two phonon branches: longitudinal and transverse 
\cite{Kleinert}, provided that the following condition is fulfilled:

\begin{eqnarray}
\pi_1(Q,\omega)=\pi_2(Q,\omega)(\mu+K)/\mu\equiv \pi(Q,\omega)\,.
\label{symms12}
\end{eqnarray}

\noindent Straightforward calculations show that imaginary parts of
$\pi_{1,2}$ merely result in a weak damping of the vibration modes. This
effect will be neglected here, and only real part $Re\pi(Q,\omega)$ that
results in softening of the vibration modes will be retained.

Diagrammatic form of the electronic (hole) polarization loop $\pi(Q,\omega)$
contributing to the phonon self-energy is
exhibited in Fig.\ref{loopa}. Here the dot is electron(hole)-lattice coupling 
vertex, the  thin lines are Green's functions $G(p,\epsilon)$ of the 1D hole 
possesing momentum $p$ along the stripe:

\begin{eqnarray}
G^R(p,\epsilon)=\displaystyle\frac{1}{\epsilon-\tilde{\epsilon}(p)+i\tau_p^{-1}};
\;
\tilde{\epsilon}(p)\equiv {\epsilon}(p)-\epsilon_F=\epsilon_F(p^2-1)
\label{greens}
\end{eqnarray}

\noindent In Fig. \ref{loopb} the wavy line is added. It represents
concomitant inelastic forward scattering event with small momentum transfer 
($q\ll Q\sim 2p_F$) due to stripe vibrations described by a propagator $g(q,\omega)$.
A weak-coupling quasi classical description of the effect of meandering 
on a hole motion along the stripe is sketched in Appendix
\ref{App1}. The superscript $R$ signifies retarded Green's function, 
$\epsilon_F$ is 1D ``Fermi energy'' (chemical potential) of the on-stripe 
mobile holes, and 
momentum $p$ is normalized by 1D Fermi-momentum $p_F$; $\tau_p$ is particle
life-time. A parabolic dispersion $\epsilon(p)$ is chosen for the mobile
on-stripe holes for simplicity. A particular expression for the stripe
vibration propagator 
$g(q,\omega)$ will be elucidated below using best fit to the experimental data 
of Pintschovius et.al. \cite{pintschovius}. 
\section{Localized vibrations in a ``haircomb'' potential}   
By their mathematical structure equations
(\ref{dynamo}) map onto a set of the two coupled Schr\"odinger equations for a 
quantum particles of ``masses'' $(\mu+K)^{-1}/2$ and $\mu^{-1}/2$ in the periodic
Dirac's haircomb potentials $\propto \pi_ip(x)$.
It is important to mention that this ``potentials'' are attractive when
$Re \pi_ip(x)<0$. Hence, the real parts of the polarization loops in
Figs.\ref{loopa},\ref{loopb} have to be
negative in order to cause splitting off a narrow band of the localized 
vibrations below the
bottom of the bare propagating phonons band (the ``softening effect''). 
Corresponding Bloch solutions in the periodic potential obey the 
following quasi periodicity conditions at the adjacent intervals along the 
$x$ - axis:

\begin{eqnarray}
&&\vec{u}(x)\equiv \{u_{x}(x),u_{y}(x)\}=
\vec{A}\exp{\{\nu x\}}+\vec{B}\exp{\{-\nu x\}};\;0\leq x\leq
b;\nonumber \\
&&\vec{u}(x)=\exp\{ikb\}\left[\vec{A}\exp{\{\nu x\}}+
\vec{B}\exp{\{-\nu x\}}\right];\;
b\leq x\leq 2b. 
\label{bloch}
\end{eqnarray}

\noindent Here quasi-momentum $k$ is parallel to the $x$-axis (i.e. 
perpendicular to the stripes direction) and spans the interval 
$[-\pi/b,\pi/b]$.
The case of real $\nu$ corresponds to vibrations localized in the vicinity of
the stripes. The two-component vectors $\vec{A}$, $\vec{B}$ are defined as:

\begin{eqnarray}
\vec{A}\equiv\{A_{x},A_{y}\};\;\vec{B}\equiv\{B_{x},B_{y}\}
\label{blochab}
\end{eqnarray}

\noindent Continuity of the solution $\vec{u}(x)$ at the point $x=b$ together
with conditions for the first derivatives following from Eqs. (\ref{dynamo}):

\begin{eqnarray}
&&(\mu+K)\left[\partial_x u_x(b+0)-\partial_xu_x(b-0)\right]-\pi_1u_x(b)=0;
\nonumber\\
&&\mu\left[\partial_x u_y(b+0)-\partial_xu_y(b-0)\right]-\pi_2u_y(b)=0,
\label{bbb}
\end{eqnarray}

\noindent
lead to a single eigenvalue equation (due to a simplifying relation 
(\ref{symms12})):

\begin{eqnarray}
\cos(kb)=\cosh(\nu b)+\frac{\pi(Q,\omega)}{2\nu(\mu+K)}\sinh(\nu b);
\label{dets1}
\end{eqnarray}

\noindent This equation determines spectrum of the wave vector 
$\nu(k)$.

Substitution of the Bloch-wave (\ref{bloch}), taken in the interval
  $0<x<b$, into Eqs. (\ref{dynamo}) leads to the following system of algebraic
homogeneous equations:

\begin{eqnarray}
\left\{\begin{array}{c}
\left[\rho\omega^2-\mu Q^2+(\mu+K)\nu^2\right]A_x+iK\nu Q A_y=0;\\
iK\nu Q A_x+\left[\rho\omega^2-(\mu+K) Q^2+\mu\nu^2\right]A_y=0.
\end{array}\right.
\label{syst}
\end{eqnarray}

\noindent A dispersion $\omega(Q,k)$ is found by solving determinant 
equation of the system (\ref{syst}):

\begin{eqnarray}
 \left(\rho\omega^2-\mu
(Q^2-\nu^2)\right)\left(\rho\omega^2-(\mu+K)(Q^2-\nu^2)\right)=0
\label{dispers}
\end{eqnarray}

\noindent Solutions of (\ref{dispers}) with $\nu=\nu(k)$ result in two 
branches of dispersion $\omega_{t,L}(Q,k)$:

\begin{eqnarray}
\omega_{t,L}^2(Q,k)=c_{t,L}^2(Q^2-\nu^2(k));
\label{tL}
\end{eqnarray} 

\noindent Here $c_{t,L}$ are sound velocities of 2D elastic media 
\cite{Kleinert}:

\begin{eqnarray}
c_t^2=\mu/\rho;\;\;c_L^2=(\mu+K)/\rho.
\label{cis}
\end{eqnarray}

\noindent
Without electron-phonon coupling the corresponding unperturbed branches are:

\begin{eqnarray}
{\omega^{0}}_{t,L}^2(Q,k)=c_{t,L}^2(Q^2+k^2);
\label{tL0}
\end{eqnarray}

\noindent since at $\pi(Q,\omega)\rightarrow 0$ the eigen values of momentum
$\nu$, that are solutions of Eq. (\ref{dets1}), are purely imaginary: 
$\nu=ik$, 
and correspond to the propagating vibrations.
It is known from the ordinary quantum mechanics that localized states, 
i.e. in the present case vibrations $u_{x,y}(x)$
with real wave-vector $\nu$, have ``energies'' below the bottom of the
propagating band. Indeed, it is apparent from a direct comparison of 
Eqs. (\ref{tL}) and (\ref{tL0}), that for real $\nu(k)$: $\omega_{t,L}(Q,k)<
{\omega^{0}}_{t,L}(Q,k)$. This, in turn, signifies a phonon softening effect
due to on-stripe electron-phonon coupling.
Equation (\ref{dets1}) possesses real 
solutions $\nu(k)$ in the whole interval $-\pi\leq kb\leq \pi$ when
$Re\pi(Q,\omega)<0$ and $|Re\pi(Q,\omega)|b/(\mu+K)>4$. An imaginary 
solution $i\nu$ appears first time when $\nu(k)$ reaches zero. 
In this case equation (\ref{dets1}) reads:

\begin{eqnarray}
\cos(kb)=1-\frac{|\pi(Q,\omega)|b}{2(\mu+K)}.
\label{zero}
\end{eqnarray}

\noindent
This equality can be  satisfied when 
$0\leq |\pi(Q,\omega)|b/(\mu+K)\leq 4$. 
(Here and below the
imaginary part $Im\pi(Q,\omega)$, leading to weak damping of the modes, is 
neglected.)
Then, the lowest-value imaginary solutions 
$i\nu$ appear in the subinterval of $k$'s: 
$\arccos\left\{1-|Re\pi(Q,\omega)|b/(2(\mu+K))\right\}\leq kb\leq \pi$. 
They correspond to dispersion $\omega(Q,k)$ merging with propagating phonon 
band $\omega^{0}(Q,k)$. Finally, when $Re\pi(Q,\omega)>0$ all 
solutions $\nu(k)$ are imaginary.

In the low doping limit, assuming 
$|Re\pi|b/(\mu+K)\propto x_h^{-1}\gg 1$, the decay wave vector 
$\nu=\nu(k)$ spans a narrow band between the bounds $\nu_{\pm}$
when $k$ spans the interval $(-\pi/b,\pi/b)$ :

\begin{eqnarray}
\nu_{\pm}=-\displaystyle\frac{Re\pi(Q,\omega)}{2(\mu+K)}\left(1\mp
  2\exp\left\{\frac{Re\pi(Q,\omega)b}{2(\mu+K)}\right\}\right)
\label{nu}
\end{eqnarray}

Importantly, it follows from Eq. (\ref{nu}) that in the strong coupling limit 
$|Re\pi_i|b/(\mu+K)\gg 1$ the wave-vector $\nu(k)\approx$const, since in this
limit the exponentially
small second term in the right hand side of Eq. (\ref{nu}) can be neglected.
This makes softening independent of the hole doping concentration $x_h$
(involved in (\ref{nu}) only via interstripe period $b$), in
accord with the experiments in high-T$_c$ cuprates \cite{Qee,pintschovius}. 
Substituting
(\ref{nu}) into Eq. (\ref{tL}) one finds the following generalized relations: 

\begin{eqnarray}
\Delta\omega_{t,L}(Q,k)\approx-\displaystyle\frac{c_{t,L}^2}{2\omega_{t,L}^0(Q,k)}
\left(\frac{Re\pi(Q,\omega_L)}{2(\mu+K)}\right)^2.
\label{tlsoft}
\end{eqnarray}

\noindent where $\Delta\omega_{t,L}(Q,k)\equiv \omega_{t,L}(Q,k)-{\omega^0}_{t,L}(Q,k)$.
 Relations (\ref{tlsoft}) are remarkable, the
softening amount $\Delta\omega_{t,L}(Q,k)$ of the in-plane phonon modes is
expressed via the self-energy 
correction to the phonon Green's function arising from the on-stripe hole-phonon 
coupling. The self-energy is given by the polarization 
loop $\pi(Q,\omega)$, that carachterizes properties of the holes moving
along the valley of a single stripe. As will be demonstrated below, the flat
$Q$-dependence of the modes softening measured experimentally
\cite{Qee},\cite{pintschovius} leaves narrow room for the different
choices of statistics and dynamical characteristics of the mobile on-stripe 
holes. Thus, relations (\ref{tlsoft}) open an exclusive 
``keyhole''
into the nature of stripes in strongly correlated electron systems, and in
particular, in the underdoped high-T$_c$ cuprates. 

\section{Phonon self-energy: 1D holes inside static stripes}

First, consider static stripes and use the standard methods
\cite{abrikosov} to find expression for polarization loop of the 1D 
on-stripe holes (see Fig. \ref{loopa}): 

\begin{eqnarray}
&& \pi(Q,i\omega)=\frac{g^2}{2\pi^2}\int_{-\infty}^{\infty}d\epsilon\int dp
\left\{Im G^R(p,\epsilon)G^A(p+Q,\epsilon-i\omega)+\right.\nonumber\\
&&\left. Im G^R(p+Q,\epsilon)G^R(p,\epsilon+i\omega)\right\}
\tanh{\left(\frac{\epsilon}{2T}\right)};\; \omega=2\pi nT,\;n=0,\pm 1,...
\label{pi1d}
\end{eqnarray}

\noindent Here $g$ is electron-phonon coupling constant and $G^{R,A}$ are
retarded and advanced versions of the Green's function (\ref{greens}), 
$\omega$ is the Matsubara frequency.
After an analytic continuation into real axis of $\omega$ of the expression  
(\ref{pi1d}) and separation of the real part, one finds the
inverse localization length $\nu(k)\approx\nu(Q,\omega)$ 
using eqs. (\ref{nu}) and (\ref{pi1d}):

\begin{eqnarray}
&& \nu(Q,\omega)=-\gamma\frac{1}{2\pi}\int_{-\infty}^{\infty}d\epsilon\int dp
\left\{Im G^R(p,\epsilon)Re G^R(p+Q,\epsilon-\omega)+\right.\nonumber\\
&&
\left. Im G^R(p+Q,\epsilon)Re
G^R(p,\epsilon+\omega)\right\}\tanh{\left(\frac{\epsilon}{2T}\right)};\;\;
\label{nu1d}\\
&&
\gamma=\frac{g^2\hbar}{4\epsilon_F(\mu+K)p_FV_0}\sim 1.
\label{g1d}
\end{eqnarray}
 
\noindent Here $T$ is the temperature and $V_0$ is the volume of the unit
cell. All energies are expressed in units of $\epsilon_F$. As will become apparent below, a reasonable fit to experimental data can be 
obtained neglecting $T$ as compared with ${\tau_p}^{-1}$. The analytic
expression for $\nu(Q,\omega)$ derived in this limit is given in 
Eq. (\ref{pi1df}) in Appendix \ref{App2}. 

For a comparison, in the case of a  three-dimensional spherical Fermi surface 
the well known expression \cite{abrikosov} for polarization loop $\pi_{3D}(Q)$
results in a prediction for $\nu_{3D}$ following from Eq. (\ref{nu}):

\begin{eqnarray}
\nu_{3D}(Q,\omega)\propto
\left\{1+\frac{1-z^2}{2z}\displaystyle\ln\left|\frac{1+z}{1-z}\right|\right\};
\mbox{where:}\, z\equiv Q/2p_F.
\label{pi3d}
\end{eqnarray}

\noindent Substituting expressions for $\nu$ from eqs. (\ref{nu1d}) and
(\ref{pi3d}) into modes softening expression Eq. (\ref{tlsoft})
one obtains the dash-dotted and dotted curves in Fig. \ref{comparison} for
the 1D and 3D holes respectively. The filled circles in Fig.\ref{comparison} 
are obtained from the experimental data \cite{pintschovius} measured in 
La$_{1.85}$Sr$_{0.15}$CuO$_4$ high-T$_c$ cuprate sample (T$_c$$=38$K) by 
subtracting values of
$\omega_{LO}(Q)$ at zero doping from those measured at finite concentration of
doped holes. 
All the curves and data 
points in Fig. \ref{comparison} are normalized by their respective minimal 
values for each individual set. It is obvious
from Fig. \ref{comparison} that none of the so far mentioned theoretical 
curves 
fits reasonably well the experimental data. The 3D holes produce softening
effect that depends on wave-vector Q the way just opposite to the one exhibited
by the experimental curve. Simultaneously, phonon coupling to 1D holes on the 
static stripes, resulting in the self-energy (\ref{pi1d}), cannot reproduce the
measured \cite{Qee},\cite{pintschovius} shallow $Q$-dependence of the phonon 
modes softening, that takes place within a sizable interval of momenta 
comparable with $p_F$ itself.  

\section{Phonon self-energy: 1D holes inside dynamic stripes}
 
\begin{figure}[tbph]
\begin{center} 
\includegraphics[width=3.0in]{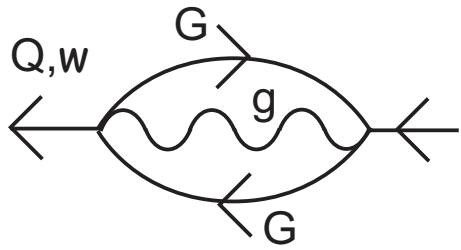}
\caption{ Same as Fig. \ref{loopa}, but with wavy line for stripe meandering 
propagator $g(q,\omega)$. }
\end{center}
\label{loopb}
\end{figure}

\begin{figure}[tbph]
\begin{center} 
\includegraphics[width=3.0in]{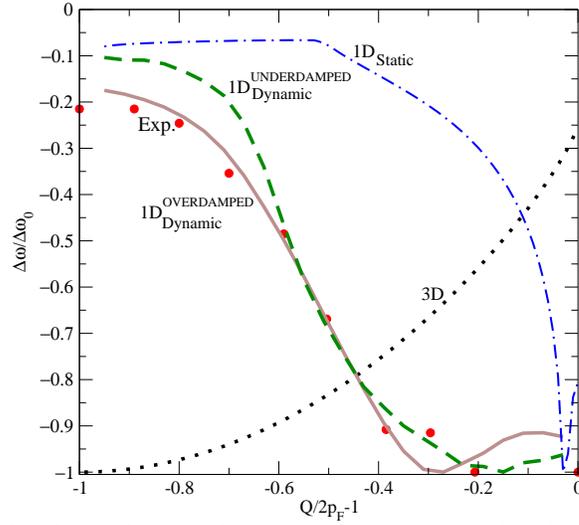}
\caption{ Softening of the phonon band. Filled circles: experimental data 
by [2] for La$_{1.85}$Sr$_{0.15}$CuO$_4$, T$_c=38K$;
solid line: 1D holes on meandering stripe, using Fig. \ref{loopb}
with wavy line (\ref{g}) in the overdamped case;
dashed line: the same as solid line, but with (\ref{g}) in  underdamped
case;
dash-dotted line: 1D holes on static stripe, Fig. \ref{loopa};
dotted line: 3D holes, using Fig. \ref{loopa} with 3D fermionic
propagators $G$. $\Delta\omega_0$
normalizes all curves at their minima to $-1$. }
\end{center}
\label{comparison}
\end{figure}

Now we demonstrate that good correspondence with experiment in underdoped 
cuprates
can be achieved by inclusion of an extra massless bosonic propagator with
small momentum transfer ($q\ll p_F$) accompanying 
on-stripe hole-phonon scattering event, as it is indicated by the wavy line in  
Fig. \ref{loopb}. This extra scattering can be ascribed to transversal stripe
meandering in the $\{x,y\}$-plane.
Expression for polarization loop 
$\Pi(Q,i\omega)$, with wavy-line included in accord with Fig. \ref{loopb}, 
looks as follows :

\begin{eqnarray}
&&Re\Pi(Q,i\omega)=-\frac{{\tilde{g}}^2}{\pi}\int_{-\infty}^{\infty}d\epsilon
\int dq
\left\{Im g^R(q,\epsilon)Re\pi^R(Q-q,\omega-\epsilon)+\right.\nonumber\\
&&\left. Im \pi^R(Q-q,\epsilon)Re g^R(q,\epsilon+\omega)\right\}
\coth{\left(\frac{\epsilon}{2T}\right)};\; \omega=2\pi nT,\;n=0,\pm 1,..\,,
\label{pi2d}
\end{eqnarray}

\noindent where momentum $q$ is expressed in the reciprocal lattice units, all
energies are expressed in units of $\epsilon_F$, and
dimensionless $\tilde{g}$ redefines the coupling constant $g$, that was 
introduced in
Eqs. (\ref{pi1d}) and (\ref{g1d}) for the case of static stripes. An explicit
expression for ${\tilde{g}}^2$ is derived below. 

Now, the choice of the ``meandering propagator'' 
$g^R(k,\omega)$ should be made. A phonon-like propagator has the form:

\begin{eqnarray} 
g^R(k,\omega)=\frac{1}{2\omega_k}\displaystyle\left(\frac{1}{\omega-\omega_k+
i\Delta}
-\frac{1}{\omega+\omega_k+i\Delta}\right),
\label{g}
\end{eqnarray} 

\noindent where $\Delta$ signifies inverse lifetime of the excitation. Below 
to cases are considered: underdamped $\omega_{k}
\propto c_{\alpha}k\gg \Delta$, and
overdamped $\omega_{k}\ll \Delta$, 
stripe meandering excitations. Before detailing the results, 
a heuristic derivation of a dimensionless $\tilde{g}$ for the four-leg vertex 
in Fig. \ref{loopb} is in order. This is most easily done using 
Eq. (\ref{wavebess}). The latter equation infers for the case
of a small
stripe meandering  $F_q$ with wave vector $q$ (along the stripe) a 
dimensionless hole-stripe 
coupling amplitude $\sim pF_q$, where $F_q\sim \sqrt{\hbar/M\omega_q}$ and 
$p\sim p_F/\hbar$. Here the property of the Bessel function: 
$J_1(x\rightarrow 0)\propto x$ at small arguments is used. The dispersion 
function $\omega_q$ enters meandering propagator $g(q,\omega)$
in Eq. (\ref{g}) and $M$ is ``mass'' associated with the stripe meandering. 
Hence, the dimensionless coupling constant $\tilde{g}$ introduced in 
Eq. (\ref{pi2d}) equals :

\begin{eqnarray}
{\tilde{g}}^2={p_F}^2/(M\epsilon_F)\sim m/M\,.
\label{gtil}
\end{eqnarray}

\noindent where $m$ is hole (electron) mass.

\subsection{Underdamped stripe meandering case}
\begin{figure}[tbph]
\begin{center} 
\includegraphics[width=3.0in]{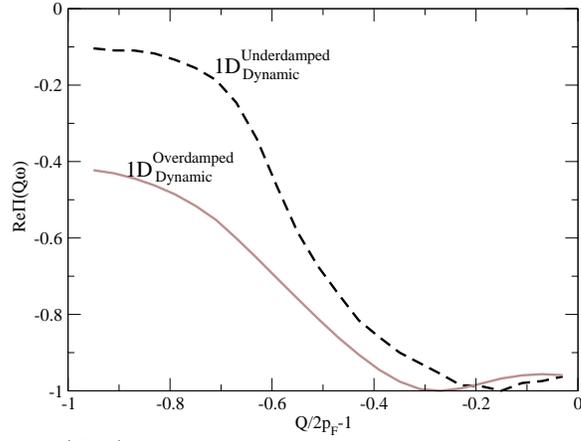}
\caption{Polarization $Re\Pi(Q,\omega)$ as function of momentum $Q$ at fixed 
$\omega=0.1\varepsilon_F$. Solid line is calculated
with overdamped phonon-like propagator Eq. (\ref{g}). Dashed
line is for underdamped case. All curves are normalized to $-1$ at their
minima.}
\end{center}
\label{loopc}
\end{figure}

First, substitute phonon-like propagator (\ref{g}) into (\ref{pi2d}), and
use results (\ref{pi1df}) and (\ref{impi1d}) for the 1D polarization loop
$\pi(Q,\omega)$. The real part 
$Re\Pi(Q,\omega)$ as function of momentum $Q$ at fixed $\omega$
(corresponding to the value $\omega_{LO}$ of the softened LO mode in cuprates) 
proves to be negative (see dashed line in Fig. \ref{loopc}), and therefore,
softening of the bare phonon band occurs according to explained in 

detail in the text after Eq. (\ref{tL0}). Substitution of the calculated
$Re\Pi(Q,\omega)$ into Eq. (\ref{tlsoft}) gives $Q$-dependence of
the mode softening. The softening value,$\Delta\omega(Q)$, normalized by its 
maximal (absolute) amount along the calculated curve, $\Delta\omega_0$, is 
plotted with dashed line in Fig. \ref{comparison}. The best fit is shown, that 
implies ${\tau_p}^{-1}/\epsilon_F=1.0$, $c_{\alpha}=0.1 v_F$ ($v_F$ is the 
Fermi velocity), 
$\Delta/\epsilon_F=0.05$, and meandering wave vector carried by the wavy line
in Fig. \ref{loopb} obeys : $|k|\leq 0.7p_F$. 
The discrepancy in the slope of the mode softening as 
function of $Q$ with the normalized experimental data \cite{pintschovius} 
(filled circles) for the LO-mode softening in La$_{1.85}$Sr$_{0.15}$CuO$_4$ 
high-T$_c$ cuprate is rather apparent. Since, according to Eq. (\ref{gtil}) : 
${\tilde{g}}^2\sim m/M\sim {c_{\alpha}}^2/{v_F}^2=0.01$ for this fit, the 
more drastic discrepancy with experiment lies in the relative mode softening 
value. It follows from relation (\ref{tlsoft}) and above estimate of 
${\tilde{g}}^2$ that :

\begin{equation}
 \Delta\omega_{t,L}/{\omega^0}_{t,L}\sim (\nu/Q)^2\approx 0.01\,,
\label{sofamu}
\end{equation}

\noindent assuming $x_h=0.15$. This is order of magnitude below
the experimental mode softening amount $\sim 0.15$ \cite{Qee,pintschovius}.
 
\subsection{Overdamped stripe meandering case}  
In this case, substitution of the calculated
$Re\Pi(Q,\omega)$ into Eq. (\ref{tlsoft}) gives $Q$-dependence that is 
plotted with solid line in Fig. \ref{comparison}. One sees impressive 
correspondence with experimental data. Again, the best fit is 
shown, that now implies ${\tau_p}^{-1}/\epsilon_F=1.0$, 
$\Delta/\epsilon_F=2.0$, 
and $|k|\leq 0.1p_F$, i.e. describing long-wavelength meandering. 
Because of the latter choice, the value of ``velocity'' $c_{\alpha}$
becomes immaterial, provided it obeys: $c_{\alpha}\lesssim 10 v_F$. Here
the relative mode softening value reaches 
$\Delta\omega_{t,L}/{\omega^0}_{t,L}\sim (\nu/Q)^2\sim 0.1$, that is 
comparable with experimental softening amount $\sim 0.15$ 
\cite{Qee,pintschovius} under condition derived using Eq. (\ref{gtil}) :

\begin{equation}
 m/M\sim 10\,.
\label{sofamu1}
\end{equation}

\section{Theory vs experiment: consequences of the correspondence}
   
The following picture emerges through the ``keyhole'' results described above.
The ``light mass'' $M$ in Eq. (\ref{sofamu1}), infiltrated in relation with 
long-wavelength stripe meandering, could be in principle understood provided 
the following conditions are fulfilled: i)the stripe meandering is very weakly 
coupled to the crystal lattice; ii) ``light mass'' $M$ has electronic origin 
and is an effective mass/per period (along stripe) of the solitonic domain 
wall separating neighboring antiphase antiferromagnetic domains of the stripe 
phase pattern \cite{Zaanen}. Next to 
mention here, is 
severe suppression of the on-stripe hole quasi-particle strength, since 
the best fit of the experimental data implies
inverse lifetime of the order of the bare Fermi-energy
$\tau^{-1}=\varepsilon_F$. The transverse meandering vibrations are also 
strongly damped, as the same fit demands that their inverse lifetime 
$\Delta=2\varepsilon_F$.

Another important consequence of the results plotted in Fig. \ref{comparison}
is the implicit value of the Fermi-momentum of the on-stripe holes: 
$p_F=\pi/2a$, where $a$ is the unit cell spacing along the stripe. Only with 
this choice the theoretical curve could be successfully superposed on the 
experimental data \cite{Qee},\cite{pintschovius}. The reason is that $Q=2p_F$ 
defines position of zero for the theoretical curves along $Q$-axis in 
Fig.\ref{comparison}. On the other 
hand, a weak-coupling consideration using spinful fermionic holes
that fill half of the on-stripe sites, leads
to the estimate : $p_F=\pi/4a$ \cite{shen}. The discrepancy could be 
tolerated 
if one assumes that on-stripe particles interacting with LO phonons are
spinless fermions, since in this case their Fermi-momentum $p_F^s$ is twice
the spinful one: $p_F^s=2p_F$. Then, one has to assume that the ARPES 
experiments in Nd-doped La$_{1.28}$Nd$_{0.6}$Sr$_{0.12}$CuO$_4$ \cite{shen}, 
probably, register already 
``spin-dressed''
(spinful) fermions, since the ``holly cross'' pattern of the occupied 1D
electron states gives
$2p_F=\pi/2a$ for the distance between parallel pieces of the quasi 1D 
``Fermi surface''. The latter is 
ascribed to 1D motion of holes along array of the parallel stripes.

\section*{ACKNOWLEDGEMENTS}
The author acknowledges important communications with Alan Bishop and 
Jan Zaanen concerning facts related with phonon softening and concepts 

of dynamic stripe-phase in high-T$_c$ cuprates.

\appendix
 \section{Phonon self-energy calculations}
\label{App2}
After the substitution 
$\tanh(\epsilon/2T)\rightarrow \mbox{sign}\epsilon$, a direct integration over 
variable $\epsilon$ in the expression (\ref{nu1d}) leads to the following answer:

\begin{eqnarray}
&&\nu(Q,\omega)=\gamma\frac{1}{2\pi}\int dp\left\{\frac{1}{\tilde{\epsilon}_1-
\tilde{\epsilon}_2-\omega}\left[\arctan\left(\frac{\tau^{-1}}{\tilde{\epsilon}_2+
\omega}
\right)-\arctan\left(\frac{\tau^{-1}}{\tilde{\epsilon}_1}\right)-\right.\right.
\nonumber\\
&&\left.\left. \arctan\left(\frac{\tau^{-1}}{\tilde{\epsilon}_1-\omega}\right)+
\arctan\left(\frac{\tau^{-1}}{\tilde{\epsilon}_2}\right)\right]-
\displaystyle\frac{1}{(\tilde{\epsilon}_1-\tilde{\epsilon}_2-\omega)^2+
4\tau^{-2}}\displaystyle\left[\left(\tilde{\epsilon}_1-\tilde{\epsilon}_2-\omega\right)
\times\right.\right. \nonumber\\
&&\left.\left[\arctan\left(\frac{\tau^{-1}}{\tilde{\epsilon}_2+\omega}
\right)+\arctan\left(\frac{\tau^{-1}}{\tilde{\epsilon}_1}\right)-
\arctan\left(\frac{\tau^{-1}}{\tilde{\epsilon}_1-\omega}\right)-
\arctan\left(\frac{\tau^{-1}}{\tilde{\epsilon}_2}\right)
\right]+\right.\nonumber\\
&&\left.\left.\tau^{-1}\ln\left(\frac{(\tilde{\epsilon}_2+\omega)^2+\tau^{-2}}
{{\tilde{\epsilon}_1}^2+\tau^{-2}}\frac{(\tilde{\epsilon}_1-\omega)^2+\tau^{-2}}
{{\tilde{\epsilon}_2}^2+\tau^{-2}}\right)\right]\right\}\label{pi1df}.
\end{eqnarray}

\noindent In accord with the formula (\ref{nu}) for the wave vector $\nu$  
expression (\ref{pi1df}) gives (modulo constant prefactor) the real part
of the polarization loop, $Re \pi^R(Q,\omega)$. Simultaneously, the imaginary 
part $Im \pi^R(Q,\omega)$, derived from
Eq. (\ref{pi1d}), equals:

\begin{eqnarray}  
&&Im \pi^R(Q,\omega)=\frac{g^2}{\varepsilon_FV_0}\frac{1}{2\pi}\int dp
\left\{\frac{\tau^{-1}}
{(\tilde{\epsilon}_1-\tilde{\epsilon}_2-\omega)^2+4\tau^{-2}}
\left[\arctan\left(\frac{\tau^{-1}}{\tilde{\epsilon}_2+
\omega}\right)+\right.\right.
\nonumber\\
&&\left.\left.\arctan\left(\frac{\tau^{-1}}{\tilde{\epsilon}_1}\right)- 
\arctan\left(\frac{\tau^{-1}}{\tilde{\epsilon}_1-\omega}\right)-
\arctan\left(\frac{\tau^{-1}}{\tilde{\epsilon}_2}\right)\right]+
\right.\nonumber\\
&&\left.\displaystyle\displaystyle\frac{\tau^{-2}}{(\tilde{\epsilon}_1-
\tilde{\epsilon}_2-\omega)
[(\tilde{\epsilon}_1-\tilde{\epsilon}_2-\omega)^2+4\tau^{-2}]}
\left[\ln\left(\frac{(\tilde{\epsilon}_2+\omega)^2+\tau^{-2}}
{{\tilde{\epsilon}_1}^2+\tau^{-2}}\right)+\right.\right.
\nonumber\\
&&\left.\left. \ln\left(\frac{(\tilde{\epsilon}_1-\omega)^2+
\tau^{-2}}
{{\tilde{\epsilon}_2}^2+\tau^{-2}}\right)\right]\right\},
 \label{impi1d}
\end{eqnarray}

\noindent The following choice of $\arctan(x)$ branches is adopted:

\begin{eqnarray}
\arctan(x\rightarrow\pm\infty)=\frac{\pi}{2};\;\arctan(x\rightarrow
0^{-})=\pi;\;\arctan(x\rightarrow 0^{+})=0.
\end{eqnarray}

\noindent Here: $\tilde{\epsilon}_1\equiv \tilde{\epsilon}(p)$ and 
$\tilde{\epsilon}_2\equiv \tilde{\epsilon}(p+Q)$, and all energies and inverse 
life-times are expressed in units of the bare Fermi-energy $\varepsilon_F$ of
the on-stripe holes. The difference between $\tau_{p}$ and $\tau_{p+Q}$ is
neglected.

\section{Stripe meandering effect on 1D hole motion}
\label{App1}

Consider a
simple on-stripe ``plane wave'' function $\Psi_p(x)$ of electron
(hole) moving along a stripe (i.e. dynamically curved 1D space), that is
preferentially oriented parallel to the $x$-axis. 
The actual length $L(x,t)$ of the dynamically curved stripe equals
(see e.g. \cite{Kivelson}):

\begin{eqnarray}
L(x,t)=\int_0^x dx' \displaystyle\sqrt{1+(\partial_{x'}Y(x',t))^2}+L(0),
\label{wiggle}
\end{eqnarray}

\noindent where $Y(x,t)$ describes space($x$) and time($t$)-dependent deviation of the
stripe from the straight line. Assuming for
simplicity that
$(\partial_{x'}Y(x',t))^2\ll 1$ and making expansion of the square root one finds:

\begin{eqnarray}
&&\Psi_p(x)=\exp(ipL(x,t))\approx\exp\{ip(L(0)+x)+ipF(x,t)\}; \;\;\label{wavefun}\\
&&F(x,t)=\frac{1}{2}
\int_0^x dx' (\partial_{x'}Y(x',t))^2=F_0(t)x+\sum_qF_q\sin{(qx-\omega_q
  t+\theta_q)} \label{harms}
\end{eqnarray}

\noindent In (\ref{harms}) the stripe-line length is expanded over a
harmonics, each one being characterized with wave vector $q$, frequency
$\omega_q$, and phase $\theta_q$  of the corresponding transverse vibration mode of the
stripe. Considering separately effect on the wave-function (\ref{wavefun}) of each
particular $q$-term from the sum in (\ref{harms}), it is convenient to use
representation of 
$\exp\{ipF\sin(x)\}$ via Bessel functions $J_n$ of integer order $n$ :

\begin{eqnarray}
\Psi_p(x)\approx\exp\{ip[L(0)+x(1+F_0(t))]\}\prod_q\sum_n
J_n(pF_q)\exp\{in(qx-\omega_qt+\theta_q)\}
\label{wavebess}
\end{eqnarray}

\noindent Hence, according to Eq.(\ref{wavebess}) a long-wavelength vibration
of the stripe causes (forward) scattering
$p\rightarrow p+nq$ of the on-stripe holes and $J_n(pF_q)$ gives its amplitude
(scattering matrix element).

\newpage

\newpage

\newpage


\end{document}